\documentclass[showpacs,aps,prd,preprint,nofootinbib,showkeys,unsortedaddress,raggedbottom]{revtex4-1}
\pdfoutput=1

\usepackage{bm}
\usepackage{amsmath}
\usepackage{graphicx}
\usepackage{subfigure}
\usepackage{float}
\usepackage[usenames,dvipsnames]{color}
\definecolor{darkblue}{RGB}{0,0,196}
\usepackage[colorlinks=true,linkcolor=darkblue,citecolor=darkblue,urlcolor=darkblue]{hyperref}

\usepackage{setspace}
\usepackage{footmisc}
\usepackage[makeroom]{cancel}

\newcommand{\intdP}{\int\!dP}

\def\be{\begin{equation}}
\def\ee{\end{equation}}
\def\ba{\begin{eqnarray}}
\def\ea{\end{eqnarray}}

\begin{document}

\title{Anisotropic hydrodynamic modeling of 200 GeV Au-Au collisions}

\author{Dekrayat Almaalol}
\affiliation{Department of Physics, Kent State University, Kent, OH 44242 United States}

\author{Mubarak Alqahtani} 
\affiliation{Imam Abdulrahman Bin Faisal University, Dammam 34212, Saudi Arabia}

\author{Michael Strickland} 
\affiliation{Department of Physics, Kent State University, Kent, OH 44242 United States}

\begin{abstract}
We compare phenomenological results from 3+1d quasiparticle anisotropic hydrodynamics (aHydroQP)  with experimental data collected in RHIC 200 GeV Au-Au collisions. We present comparisons of identified particle spectra in different centrality clases, charged particle multiplicity versus pseudorapidity, identified particle multiplicity versus centrality across a wide range of particle species, identified particle elliptic flow versus transverse momentum, and charged particle elliptic flow as a function of transverse momentum and rapidity.  We use the same aHydroQP and hadronic production/feed down codes that were used previously to describe LHC 2.76 TeV data.  The aHydroQP hydrodynamic model includes the effects of both shear and bulk viscosities in addition to an infinite number of transport coefficients computed self-consistently in the relaxation time approximation.   To convert to the final state hadrons, we use anisotropic Cooper-Frye freeze-out performed on a fixed-energy-density hypersurface and compute the production/feed down using a customized version of Therminator 2. We find good agreement with many heavy-ion collision observables using only smooth Glauber initial conditions parameterized by an initial central temperature of $T_0 = 455$ MeV, a constant shear viscosity to entropy density ratio $\eta/s= 0.179$, and a switching (freeze-out) temperature of $T_{\rm FO}=130$ MeV.
\end{abstract}

\date{\today}



\maketitle

\section{Introduction}
\label{sec:intro}

Ultra-relativistic heavy-ion collisions (URHICs) performed at the Relativistic Heavy Ion Collider (RHIC) at Brookhaven National Laboratory and the Large Hadron Collider (LHC) at CERN aim to create and study the properties of a deconfined quark-gluon plasma (QGP).  At zero baryochemical potential, the QGP is expected to be generated when temperatures exceed approximately 155 MeV.  In the QGP phase, quarks and gluons, which are liberated from incoming nuclei, are the relevant degrees of freedom.  In the high-energy limit, color transparency results in the central rapidity region of the matter generated in such collisions having net baryon number close to zero which mimics conditions generated in the early universe, however, the lifetime of the QGP created in URHICs is on the order of 10 fm/c, which begs the question of whether or not the matter generated can be described using models which assume that the system is in local isotropic thermal equilibrium.  The ability of ideal hydrodynamics to qualitatively describe soft hadron production and flow  \cite{Huovinen:2001cy,Hirano:2002ds,Kolb:2003dz} led to early claims that the QGP was isotropic and thermal on a time scale on the order of 0.5 fm/c (see e.g. \cite{Kolb:1999it,Heinz:2001ax,Heinz:2002un,Jacobs:2004qv,Muller:2005en,Strickland:2007fm}).

We now know that this claim was premature and that the QGP generated in URHICs is most likely momentum-space anisotropic in the local rest frame (LRF) of the matter, with the anisotropies being largest at early times $\tau \lesssim 2$ fm/c and near the transverse/longitudinal edges of the QGP at all times \cite{Ryblewski:2013jsa,Strickland:2014pga,Alqahtani:2017mhy}.  Traditionally, the existence of LRF momentum-space anisotropies is accounted for in the context of second-order viscous hydrodynamics by formally expanding in a gradient expansion around a locally isotropic and thermal state.  Early work along these lines was presented in a series of papers by Mueller, Israel, and Stewart (MIS) decades ago \cite{Muller:1967zza,Israel:1976tn,Israel:1979wp}.  The success of viscous relativistic hydrodynamics resulted in many works which have improved upon the MIS formalism and the resulting improved second-order hydrodynamical models have had great phenomenological success in describing a host of URHIC data \cite{Muronga:2001zk,Muronga:2003ta,Muronga:2004sf,Heinz:2005bw,Baier:2006um,Romatschke:2007mq,Baier:2007ix,Dusling:2007gi,Luzum:2008cw,Song:2008hj,Heinz:2009xj,El:2009vj,PeraltaRamos:2009kg,PeraltaRamos:2010je,Denicol:2010tr,Denicol:2010xn,Schenke:2010rr,Schenke:2011tv,Bozek:2011wa,Niemi:2011ix,Denicol:2011fa,Niemi:2012ry,Bozek:2012qs,Denicol:2012cn,Denicol:2012es,PeraltaRamos:2012xk,Jaiswal:2013npa,Jaiswal:2013vta,Calzetta:2014hra,Denicol:2014vaa,Denicol:2014mca,Jaiswal:2014isa}.  Currently, there is a concerted effort to quantitatively extract QGP properties such as the average initial central temperature of the QGP, its shear viscosity, its bulk viscosity, jet energy loss, heavy quark momentum diffusion constant, etc.  For recent reviews of relativistic hydrodynamics in the context of the QGP, we refer the reader to Refs.~\cite{Jeon:2016uym,Romatschke:2017ejr,Florkowski:2017olj}.

In parallel to these developments in second-order viscous hydrodynamics theory and phenomenology, there have been theoretical and phenomenological works dedicated to relaxing the assumption that the QGP is close to local isotropic thermal equilibrium in order to better account for large deviations from isotropy. To address this issue, a framework called anisotropic hydrodynamics (aHydro) was introduced~\cite{Florkowski:2010cf,Martinez:2010sc}.  This framework allows one to describe a system that is far from equilibrium (isotropy) without breaking important physics constraints such as the positivity of the one-particle distribution function, etc.  In all cases where exact solutions to the Boltzmann equation are available, it has been found that aHydroQP provides a much better approximation to the exact solutions than standard second-order schemes, particularly when the system is far from equilibrium \cite{Florkowski:2013lza,Florkowski:2013lya,Florkowski:2014sfa,Nopoush:2014pfa,Denicol:2014xca,Denicol:2014tha,Heinz:2015gka,Molnar:2016gwq,Strickland:2017kux}.

There has been a significant body of work produced since the two early aHydro papers  \cite{Florkowski:2010cf,Martinez:2010sc}, which has extended the aHydro formalism to full 3+1d dynamics, self-consistent inclusion of non-conformality of the QGP using a quasiparticle framework, etc. \cite{Ryblewski:2010ch,Florkowski:2011jg,Martinez:2012tu,Ryblewski:2012rr,Bazow:2013ifa,Tinti:2013vba,Nopoush:2014pfa,Tinti:2015xwa,Bazow:2015cha,Strickland:2015utc,Alqahtani:2015qja,Molnar:2016vvu,Molnar:2016gwq,Alqahtani:2016rth,Bluhm:2015raa,Bluhm:2015bzi,Alqahtani:2017jwl,Alqahtani:2017tnq,Martinez:2017ibh,McNelis:2018jho}.  For a recent aHydro review, we refer the reader to Ref.~\cite{Alqahtani:2017mhy}.  In this paper, we will use quasiparticle aHydro (aHydroQP) which implements the QCD equation of state via a temperature dependent quasiparticle mass that is fit to lattice data for the entropy density \cite{Alqahtani:2015qja}.  The aHydroQP formalism has been previously applied to LHC 2.76 TeV collisions and it was demonstrated that one could reproduce the observed identified particle differential spectra, charged particle multiplicity, elliptic flow, and Hanbury-Brown-Twiss radii~\cite{Alqahtani:2017jwl,Alqahtani:2017tnq}.  Using smooth Glauber initial conditions, the extracted initial central temperature was \mbox{$T_0 = 600$} MeV at $\tau_0 = 0.25$ fm/c and the extracted shear viscosity to entropy density ratio was $\eta/s = 0.159$ with a switching (freeze-out) temperature of $T_{\rm FO} = 130$ MeV.  

In this paper, we present the corresponding determination at RHIC's highest collision energy.  Keeping the switching temperature fixed, since this should be independent of the collision energy, we present fits to spectra, multiplicities, etc. and extract values of $\eta/s$ and $T_0$ for 200 GeV Au-Au collisions.  Our best fits to the spectra resulted in \mbox{$T_0  = 455$ MeV} at $\tau_0 = 0.25$ fm/c and \mbox{$\eta/s = 0.179$}.  With these values, we find good agreement with the available data for the identified particle spectra, elliptic flow in non-central centrality classes, charged particle multiplicities, and identified particle multiplicities across a broad range of centrality classes.  The resulting parameter set for the highest RHIC energies will be useful in other contexts where aHydro is being used for background evolution, e.g. for bottomonium suppression calculations \cite{Strickland:2011mw,Strickland:2011aa,Krouppa:2015yoa,Krouppa:2016jcl,Krouppa:2017lsw,Krouppa:2017jlg}.  In addition, this study will serve as a baseline for future work including, e.g., varying the initial momentum-space anisotropy, fluctuating initial conditions, finite density, etc.

The structure of the paper is as follows.  In Sec.~\ref{sec:setup} we present our setup and background for the aHydroQP calculations.  In Sec.~\ref{sec:results} we present our main results and compare to experimental data from PHENIX, PHOBOS, and STAR collaborations.  In Sec.~\ref{sec:conclusions} we summarize our findings and present an outlook for the future.

\section*{Conventions and notation}

The Minkowski metric tensor is taken to be ``mostly minus'', i.e. $g^{\mu\nu}={\rm diag}(+,-,-,-)$. The transverse projection operator $\Delta^{\mu\nu} = g^{\mu\nu}{-}u^\mu u^\nu$ is used to project four-vectors and/or tensors into the space orthogonal to $u^\mu$.   The Lorentz-invariant integration measure is \mbox{$dP = \frac{d^3{\bf p}}{(2\pi)^3} (p \cdot u)^{-1}$}.  

\section{Setup and background}
\label{sec:setup}

In anisotropic hydrodynamics, we take the one-particle distribution function to be of generalized Romatschke-Strickland form \cite{Romatschke:2003ms,Romatschke:2004jh}
\be
f(x,p) = f_{\rm eq}\!\left(\frac{1}{\lambda}\sqrt{p_\mu \Xi^{\mu\nu} p_\nu}\right) ,
\label{eq:genf}
\ee
where $\lambda$ is a temperature-like scale.  One can decompose the symmetric anisotropy four-tensor $\Xi^{\mu\nu} = u^\mu u^\nu + \xi^{\mu\nu} - \Delta^{\mu\nu}\Phi$, with $\Phi$ encoding bulk viscous corrections, $u_\mu \xi^{\mu\nu} =u_\nu  \xi^{\mu\nu} = 0$ and ${\xi^\mu}_\mu = 0$ \cite{Nopoush:2014pfa}.  Taking the anisotropy tensor $\xi^{\mu\nu}$ to be diagonal, one can rewrite the distribution function in the LRF in terms of the spacelike diagonal anisotropy factors, $\alpha_i $, 
\be 
f(x,p)= f_{\rm eq}\!\left(\frac{1}{\lambda}\sqrt{\sum_i  \frac{p_i^2}{\alpha_i^2}+m^2}\right) ,
\label{eq:genf2}
\ee
with $ i \in \{x,y,z\}$ and $\alpha_i \equiv (1+ \xi_i +\Phi)^{-1/2}$.  During the evolution phase, $m$ is a function of the local temperature and, for the purposes of freeze-out, $m$ is the mass of the particular hadron species under consideration.

\subsection{The quasiparticle Boltzmann equation}

For a system of quasiparticles with a temperature-dependent mass $m(T)$, the Boltzmann equation is \cite{Jeon:1995zm,Romatschke:2011qp,Alqahtani:2015qja}
\be
p^\mu\partial_\mu f+\frac{1}{2}\partial_i m^2\partial^i_{(p)}f =-{\cal C}[f]\, .
\ee
The quasiparticle mass $m(T)$ is determined uniquely from the lattice entropy density.  In order to conserve energy and guarantee thermodynamic consistency it is necessary to introduce a temperature-dependent background contribution $B(T)$ which is determined uniquely in terms of $m(T)$ and a boundary condition at $T=0$ using
\be
\partial_\mu B = -\frac{1}{2} \partial_\mu m^2 \intdP \,  f(x,p)\,.
\ee
For the details of the equation of state implementation, the resulting temperature-dependent quasiparticle mass $m(T)$ and background contribution $B(T)$, and the associated bulk viscosity, we direct the reader to Refs.~\cite{Alqahtani:2015qja,Alqahtani:2017tnq}.

For the collisional kernel, we use the RTA kernel
\be
C[f] = \frac{p\cdot u}{\tau_{\rm eq}} [ f - f_{\rm eq}(T) ] \, .
\ee
The relaxation time for massive quasiparticles is
\be 
\tau_{\rm eq}(T)= \bar{\eta} \, \frac{{\cal E+P}}{I_{3,2}(\hat{m}_{\rm eq})} \, ,
\ee
where $\bar\eta = \eta/s$ and
\ba 
I_{3,2}(x) &=& \frac{N_{\rm dof} T^5\, x^5}{30 \pi^2} \bigg[ \frac{1}{16} \Big(K_5(x)-7K_3(x)+22 K_1(x) \Big)-K_{i,1}(x) \bigg]  \, , \nonumber \\ 
K_{i,1}(x)&=&\frac{\pi}{2}\Big[1-x K_0(x) {\cal S}_{-1}(x)-xK_1(x) {\cal S}_0(x)\Big] \, ,
\ea
where $N_{\rm dof}$ is the number of degrees of freedom, $K_n$ are modified Bessel functions of the second kind, and ${\cal S}_n$ are modified Struve functions \cite{Alqahtani:2015qja}.

\subsection{The energy-momentum tensor}

The background contribution $B(T)$ is added to the kinetic energy-momentum tensor as follows
\be
T^{\mu\nu}=T^{\mu\nu}_{\rm kinetic}+B(T) g^{\mu\nu} \, ,
\ee
and, assuming a diagonal anisotropy tensor in the LRF, we can expand the energy-momentum tensor as
\be
T^{\mu\nu}={\cal E}u^\mu u^\nu+{\cal P}_x X^\mu X^\nu+{\cal P}_y Y^\mu Y^\nu+{\cal P}_z Z^\mu Z^\nu \, .
\label{eq:T-expan}
\ee
where $u^\mu$ is the timelike fluid velocity and $X^\mu$, $Y^\mu$, and $Z^\mu$ are spacelike four-vectors which span the directions orthogonal to $u^\mu$.\footnote{Details concerning the vector basis used can be found in Refs.~\cite{Alqahtani:2015qja,Alqahtani:2017tnq}.}  Below we will also use a compact notation \mbox{$X^\mu_i = (X^\mu,Y^\mu,Z^\mu)$}.  The energy density and pressures appearing in Eq.~\eqref{eq:T-expan} each include kinetic and background field contributions with $B(T)$ adding to the kinetic energy density and subtracting from each of the kinetic pressures isotropically.

\subsection{Equations of motion}

To obtain the equations of motion, we take moments of the Boltzmann equation using an integral operator of the form $\hat{\cal O}^{\mu\nu \cdots \lambda} = \int dP \, p^\mu p^\nu \cdots   p^\lambda$.

\subsubsection{First moment of the Boltzmann equation}

The first moment of the Boltzmann equation provides four equations
\ba
D_u{\cal E}+{\cal E}\theta_u + \sum_i {\cal P}_i u_\mu D_i X^\mu_i &=& 0 \, , \nonumber \\
D_i {\cal P}_i+{\cal P}_i\theta_i -{\cal E}X_\mu^i D_uu^\mu - \sum_j {\cal P}_j X_\mu^i D_j X^\mu_j &=& 0 \, , 
\ea
where $ i \in \{x,y,z\}$ and $j = \{x,y,z\}$\textbackslash$\{i\}$.  For example, for $i=x$, $j \in \{y,z\}$.  Above, $D_u$ is the comoving derivative along the direction of the fluid velocity, $D_u = u^\mu \partial_\mu$.  Likewise, $D_i = X^\mu_i \partial_\mu$.

\subsubsection{Second moment of the Boltzmann equation}

For the second moment we encounter the rank three tensor
\be
{\cal I}^{\mu\nu\lambda} \equiv \intdP \, p^\mu p^\nu p^\lambda  f(x,p) \, ,
\label{eq:I-int}
\ee
which can be expanded in the vector basis as
\ba
{\cal I}^{\mu\nu\lambda} &=&\, {\cal I}_u u^\mu u^\nu u^\lambda +
\sum_i {\cal I}_i \left[ u^\mu X_i^\nu X_i^\lambda + X_i^\mu u^\nu X_i^\lambda + X_i^\mu X_i^\nu u^\lambda \right] ,
 \label{eq:Theta}
\ea
with $ i \in \{x,y,z\}$  and
\ba
{\cal I}_i &=& \alpha \, \alpha_i^2 \, {\cal I}_{\rm eq}(\lambda,m) \, , \nonumber \\ 
{\cal I}_{\rm eq}(\lambda,m) &=&  4 \pi {\tilde N} \lambda^5 \hat{m}^3 K_3(\hat{m}) \, ,
\ea
where $\alpha = \prod_i \alpha_i$, $\hat{m} = m/\lambda$, and ${\tilde N} = N_{\rm dof}/(2\pi)^3$  \cite{Nopoush:2014pfa}.

We obtain three equations of motion from the diagonal projections of the second moment of the Boltzmann equation using $X_\mu X_\nu \partial_\alpha {\cal I}^{\alpha \mu \nu}$, $Y_\mu Y_\nu \partial_\alpha {\cal I}^{\alpha \mu \nu}$, and $Z_\mu Z_\nu \partial_\alpha {\cal I}^{\alpha \mu \nu}$ giving~\cite{Alqahtani:2015qja}
\be
D_u {\cal I}_i + {\cal I}_i (\theta_u + 2 u_\mu D_i X^\mu_i)
= \frac{1}{\tau_{\rm eq}} \Big[ {\cal I}_{\rm eq}(T,m) - {\cal I}_i \Big] , \label{eq:2ndmoment} 
\ee
where, once again, $ i \in \{x,y,z\}$.

\subsubsection{The effective temperature}

In the above equations we have eight unknowns $u_i$, $\alpha_i$, $\lambda$, and $T$ and seven equations (four from the first moment and three from the second moment).  To close the equations we obtain the effective temperature $T$ from the non-equilibrium energy density, ${\cal E}({\boldsymbol\alpha},\lambda,m) = {\cal E}_{\rm eq}(T)$.  This is referred to as the matching condition.  In RTA, this matching condition is a direct consequence of enforcing energy-momentum conservation.  The resulting matching condition is
\be
{\cal H}_3({\boldsymbol\alpha},\hat{m}) \lambda^4 = {\cal H}_{3,\rm eq}(1,\hat{m}_{\rm eq}) T^4 \, ,
 \label{eq:matching}
\ee
with $\hat{m}_{\rm eq} = m/T$.  The function ${\cal H}_3$ and its efficient evaluation are described in Appendix B of Ref.~\cite{Alqahtani:2017tnq}.

\subsection{Freeze-out prescription}

To convert from a fluid description to a particle description, we use the same form for the one-particle distribution function \eqref{eq:genf2} as used in the evolution.  We first extract a fixed energy density hypersurface corresponding to an effective temperature of $T_{\rm FO} = 130$ MeV and extract the microscopic parameters ($\boldsymbol{\alpha}_i, \lambda, {\bf u}$) on the hypersurface.  We then loop over 370 hadrons and hadron resonances indexed by $i$ and integrate over the hypersurface
\begin{equation}
\bigg(p^0\frac{dN}{d^3p}\bigg)_{i}=\frac{{\cal N}_i}{(2\pi)^3}\int \! f_i(x,p) \, p^\mu d^3\Sigma_\mu \, ,
\label{eq:dNdp3}
\end{equation}
to extract the hadronic spectra.  Above, the mass in the distribution function is the mass of the $i^{\rm th}$ hadron and ${\cal N}_i$ counts the number of internal degrees of freedom (spin, etc.) for hadron species $i$.  For details on the hypersurface parameterization used and other details, we refer the reader to Sec.~VI of Ref.~\cite{Nopoush:2015yga}.  The resulting parameterized hypersurface and microscopic variables are then exported in a format suitable for use by Therminator 2 \cite{Chojnacki:2011hb}.  We produced a customized version of Therminator 2 which allows for an ellipsoidally-deformed distribution function on the freeze-out hypersurface.  This customized version takes care of all hadronic production, decays, and resonance feed downs.  The 3+1d aHydroQP code and the customized version of Therminator 2 are both available publicly using the URL found in Ref.~\cite{ahydrorepo}.

\section{Results}
\label{sec:results}

We now turn to our phenomenological results.  We consider $\sqrt{s_{\rm NN}} = 200$ GeV Au-Au collisions and compare to data from the PHENIX, STAR, and PHOBOS collaborations.  In all results presented herein, we used our customized version of Therminator 2 to Monte-Carlo sample production and decays based on the freeze-out hypersurface and microscopic parameters provided by aHydroQP.  Depending on the observable and the centrality class considered, we used between 10,500 and 100,500 Monte-Carlo sampled hadronic production/decay events.  In all plots, the shaded bands surrounding the theory results indicate the statistical uncertainty associated with the hadronic production/decay sampling.

\subsection{Initial conditions}

For the initial conditions, we assume the system to be initially isotropic in momentum space ($ \alpha_{i}(\tau_0)=1 $), with zero transverse flow ($ {\bf u}_{\perp}(\tau_0) =0$), and Bjorken flow in the longitudinal direction ($ \vartheta(\tau_0) = \eta $). In the transverse plane, the initial energy density distribution is obtained from a ``tilted'' Glauber profile \cite{Bozek:2010bi}.  The profile function used was a linear combination of smooth Glauber wounded-nucleon and binary-collision density profiles, with a binary-collision mixing factor of $\chi = 0.145$.  In the rapidity direction, we used a profile with a central plateau and Gaussian ``tails'' in the fragmentation region
\be
\rho(\varsigma) \equiv \exp \left[ - (\varsigma - \Delta \varsigma)^2/(2 \sigma_\varsigma^2) \, \Theta (|\varsigma| - \Delta \varsigma) \right] .
\label{eq:rhofunc}
\ee
The parameters entering Eq.~\eqref{eq:rhofunc} were fitted to the observed pseudorapidity distribution of charged hadrons with the results being $\Delta\varsigma = 1.4$ and $\sigma_{\varsigma} = 1.4$.  The variable $\Delta\varsigma$ sets the initial width of the central plateau and the variable $\sigma_{\varsigma}$ sets the initial width of the fragmentation Gaussians.  

The resulting initial energy density at a given transverse position ${\bf x}_\perp$ and spatial rapidity $\varsigma$ was computed using 
\be 
{\cal E}({\bf x}_\perp,\varsigma) \propto (1-\chi) \rho(\varsigma) \Big[ W_A({\bf x}_\perp) g(\varsigma) + W_B({\bf x}_\perp) g(-\varsigma)\Big] + \chi \rho(\varsigma) C({\bf x}_\perp) \, ,
\ee
 where $W_{A,B}({\bf x}_\perp)$ is the wounded nucleon density for nucleus $A$ or $B$, $C({\bf x}_\perp)$ is the binary collision density, and $g(\varsigma)$ is the tilt function.  The tilt function  is defined through
\ba 
g(\varsigma) =
\left\{ \begin{array}{lcccc}
0  & \,\,\,\,\,\,\,\,\,\,\,\,\,\,\,\ & \mbox{if} & \,\,\,
& \varsigma < -y_N \, ,
 \\ (\varsigma+y_N)/(2y_N) & & \mbox{if} &
& -y_N \leq \varsigma \leq y_N \, , \\
1 & & \mbox{if} & 
& \varsigma > y_N\, ,
\end{array}\right. \,\,\,\,\,\,\,\,\,\,\,\,\,
\ea 
where $y_N = \log(2\sqrt{s_{NN}}/(m_p + m_n))$ is the nucleon momentum rapidity \cite{Bozek:2010bi}.

\begin{figure*}[t!]
\centerline{
\includegraphics[width=0.95\linewidth]{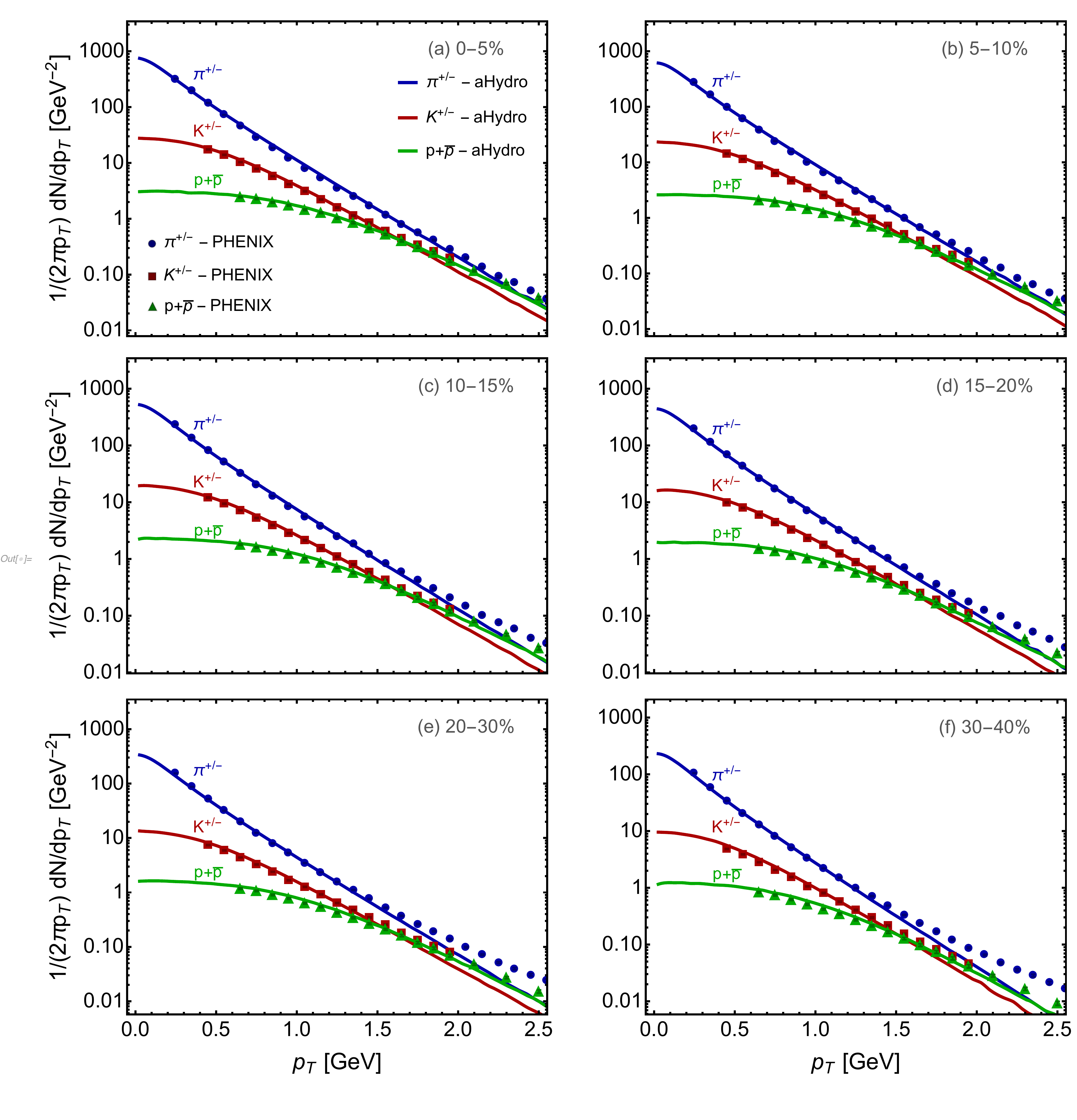}
}
\caption{Pion, kaon, and proton spectra compared to experimental observations by the PHENIX collaboration \cite{PhenixCollaboration:2004}.  The panels show the centrality classes (a) 0-5\%, (b) 5-10\%, (c) 10-15\%, (d) 15-20\%, (e) 20-30\%, and (f) 30-40\%. 
}
\label{fig:spectra-all}
\end{figure*}

\begin{figure*}[t!]
\centerline{
\includegraphics[width=.7\linewidth]{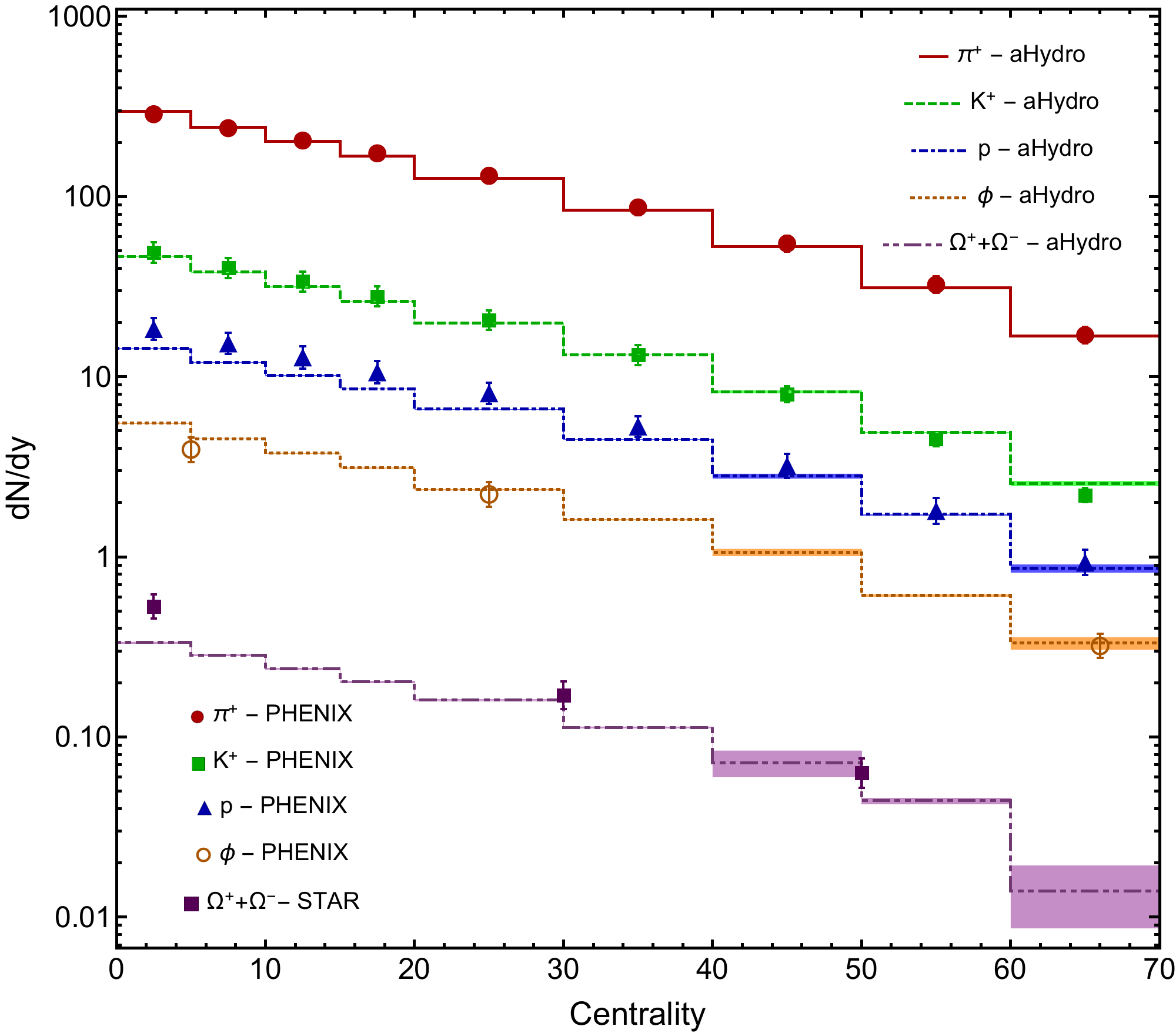}
}
\caption{Identified particle multiplicities as a function of centrality.  From top to bottom, the particles shown are $\pi^+$, $K^+$, $p$, $\phi$, and $\Omega^+ + \Omega^-$.  The data for $\pi^+$, $K^+$, and $p$ are from the PHENIX collaboration \cite{Adler:2003cb}.  Data for the $\phi$ meson production are also from the PHENIX collaboration \cite{Adams:2004ux}.  The data for $\Omega^+ + \Omega^-$ comes from the STAR collaboration \cite{Adams:2006ke}.  The aHydroQP theory results are binned using the centrality bins used by PHENIX collaboration for $\pi^+$, $K^+$, and $p$.
}
\label{fig:identified-particle-multiplicity}
\end{figure*}

\begin{figure*}[t!]
\centerline{
\includegraphics[width=.6\linewidth]{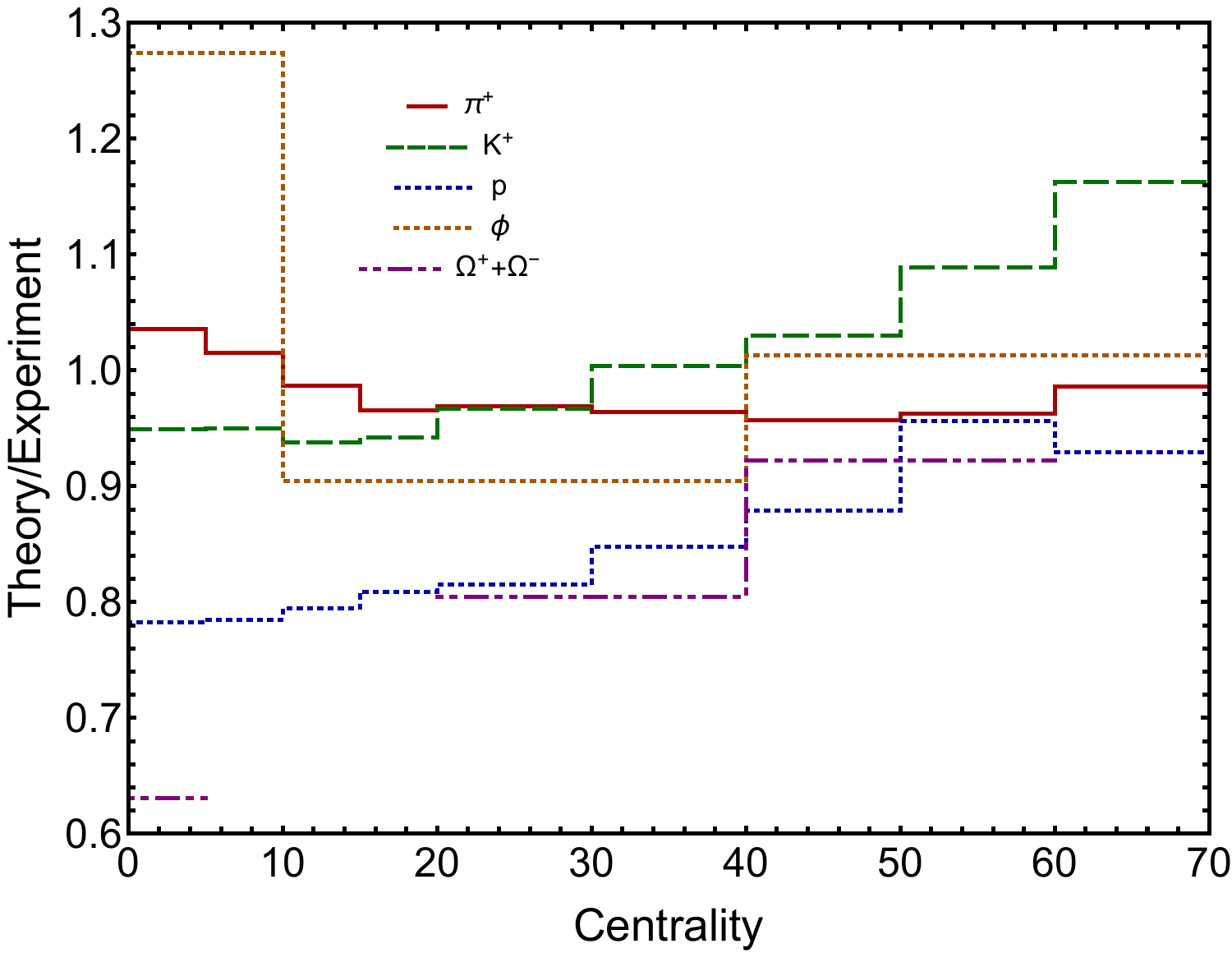}
}
\caption{Theory to data ratio for identified particle multiplicity (shown in Fig.~\ref{fig:identified-particle-multiplicity}).  For this figure, we rebinned the theory predictions for the $\phi$ and $\Omega^+ + \Omega^-$ to match with the experimental bins.}
\label{fig:identified-particle-multiplicity-error}
\end{figure*}

\subsection{Particle spectra and multiplicities}

Based on our earlier study of 2.76 TeV collisions at LHC \cite{Alqahtani:2017tnq}, we fix the switching (freeze-out) temperature to be $T_{\rm FO} = 130$ MeV.  This leaves the shear viscosity to entropy density ratio $\bar\eta = \eta/s$ and initial central temperature $T_0$ (center of the system for a $b=0$ collision) as independent parameters.  As done is our prior works, we assume that $\bar\eta$ is independent of the temperature.  In order to fit $T_0$ and $\bar\eta$, we compared model predictions with the observed pion, proton, and kaon spectra in the 0-5\% and 30-40\% centrality classes.  Based on these comparisons, we obtained \mbox{$T_0 = 455$ MeV} at $\tau_0 = 0.25$ fm/c and $\bar\eta = 0.179$.  The resulting fits to the pion, kaon, and proton spectra are shown in Fig.~\ref{fig:spectra-all} and compared to experimental data from the PHENIX collaboration \cite{PhenixCollaboration:2004}.  As can be seen from this figure, the model provides a good description of the identified particle spectra with this parameter set.  In high centrality classes, we see that the model underestimates hadron production at large transverse momentum, $p_T \gtrsim 1.5$ GeV.

In Fig.~\ref{fig:identified-particle-multiplicity} we present our results for the identified particle multiplicities as a function of centrality.  From top to bottom, the particles shown are $\pi^+$, $K^+$, $p$, $\phi$, and $\Omega^+ + \Omega^-$.  The data for $\pi^+$, $K^+$, and $p$ are from the PHENIX collaboration \cite{Adler:2003cb}.  The data for the $\phi$ meson are also from the PHENIX collaboration \cite{Adams:2004ux}.  The data for $\Omega^+ + \Omega^-$ comes from the STAR collaboration \cite{Adams:2006ke}.  The aHydroQP theory results are binned in the centrality bins used by PHENIX collaboration for $\pi^+$, $K^+$, and $p$.  As this figure demonstrates, aHydroQP coupled to our customized version of Therminator 2 is able to reproduce the centrality dependence of the observed identified particle multiplicities quite well.  This is particularly interesting because we have a single iso-thermal switching (freeze-out) temperature which is quite low, $T_{\rm FO} = 130$ MeV, and we are able to reasonably-well reproduce the observed identified particle multiplicities, not only for central collisions, but across many centrality classes.  That said, based on the theory to experiment ratios shown in Fig.~\ref{fig:identified-particle-multiplicity-error} we see that there is an approximately 20\% difference in the proton multiplicity in the 0-5\% centrality class and an approximately 35\% difference in the $\Omega^+ + \Omega^-$ multiplicity in the most central bin.  These discrepancies could have their origins  in the assumptions made during the aHydroQP evolution, freeze-out (single freeze-out with no chemical potentials), and/or the production and decays implemented by Therminator 2.

\begin{figure*}[t!]
\centerline{
\includegraphics[width=.985\linewidth]{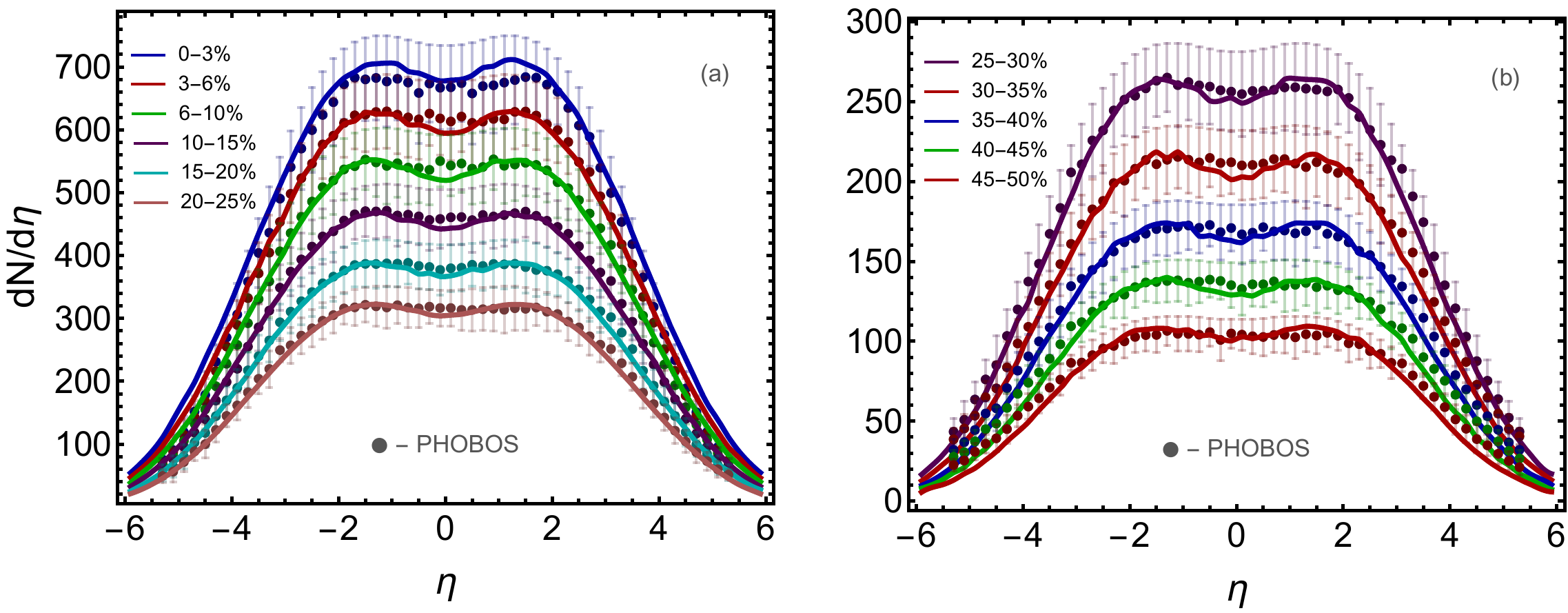}
}
\caption{aHydroQP results for charged particle multiplicity in different centrality classes (solid lines) compared to experimental data from the PHOBOS collaboration \cite{Alver:2010ck}.  Panel (a) shows centrality classes in the range 0-25\% and panel (b) shows centrality classes in the range 25-50\%.}
\label{fig:nPlotall}
\end{figure*}

In Fig.~\ref{fig:nPlotall} we present a comparison of our model predictions for the charged particle multiplicity as a function of pseudorapidity compared to experimental data from the PHOBOS collaboration \cite{Alver:2010ck}.  In panel (a) we show centrality classes in the range 0-25\% and in panel (b) we show centrality classes in the range 25-50\%.  We find that aHydroQP does a good job in reproducing the observed charged particle multiplicity as a function of pseudorapidity in a wide range of centrality classes.

Turning to collective flow, we will now present results for $v_2$.  In all cases, the theoretical result is computed using the event-plane method.  Because we use smooth Glauber initial conditions, we do not compute the higher-order harmonics.  In addition, for $v_2$ we don't expect to reproduce observations in the most centrality classes since, in such classes, elliptic flow is sensitive to event-by-event fluctuations in the geometry which are not captured by our initial conditions.  

\begin{figure*}[t!]
\centerline{
\includegraphics[width=.9\linewidth]{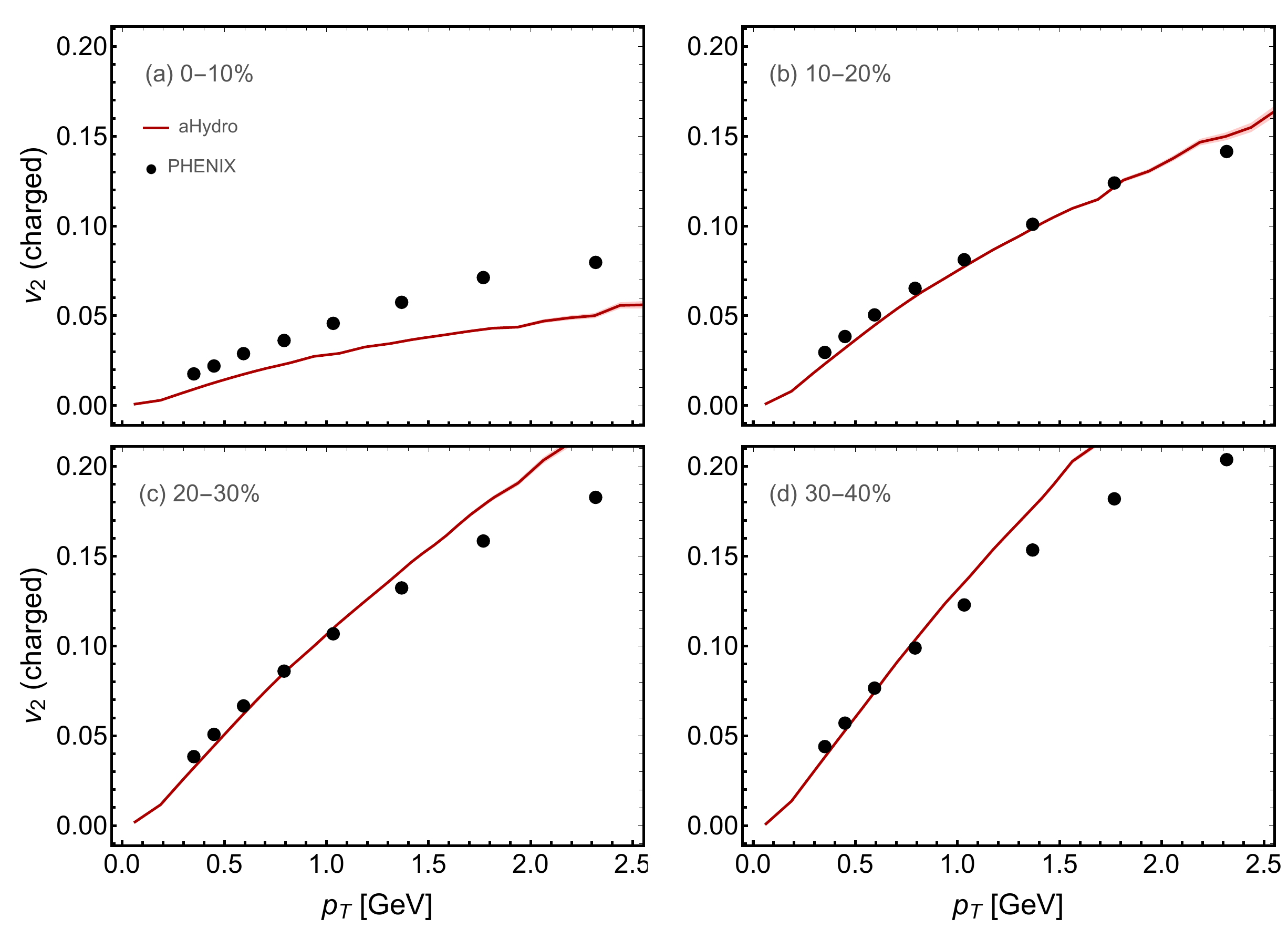}
}
\caption{Elliptic flow for charged particles in four different centrality classes:  (a) 0-10\%, (b) 10-20\%, (c) 20-30\%, and (d) 30-40\%.  The solid red line is the aHydroQP prediction and the points are observations by the PHENIX collaboration \cite{Adare:2006ti}.  The experimental errors reported by the PHENIX collaboration are smaller than the point size used.}
\label{fig:v2pt-Charged}
\end{figure*}

We start with Fig.~\ref{fig:v2pt-Charged} which shows the elliptic flow for charged particles in four different centrality classes:  (a) 0-10\%, (b) 10-20\%, (c) 20-30\%, and (d) 30-40\%.  The solid red line is the aHydroQP prediction and the points are observations by the PHENIX collaboration \cite{Adare:2006ti}.  As can be seen from this figure, aHydroQP describes the observed charged-particle $v_2$ as a function of transverse momentum reasonably well given the simplicity of the model.  We notice that in the most centrality class (0-10\%) we are underestimating the magnitude of $v_2$, but this is to be expected since we have used smooth initial conditions.  In the 20-30\% and 30-40\% centrality classes (bottom row) we see that aHydroQP over predicts $v_2$ at high transverse momentum.  In standard viscous hydrodynamics, one sees a larger ``downward bending'' of $v_2$ at high $p_T$ due to the viscous correction to the one-particle distribution function \cite{Teaney:2003kp}, however, such large corrections to one-particle distribution function call into doubt the suitability of using only terms that are linear in the viscous tensor.  In aHydroQP, there is no such truncation and terms of infinitely high order in the momentum-space anisotropies are resummed \cite{Strickland:2017kux}.  As a result, aHydroQP predicts a smaller viscous correction to the ideal hydrodynamics result for the distribution function and hence overshoots the data more than standard second-order viscous hydrodynamics treatments do.

\begin{figure*}[h!]
\centerline{
\includegraphics[width=.925\linewidth]{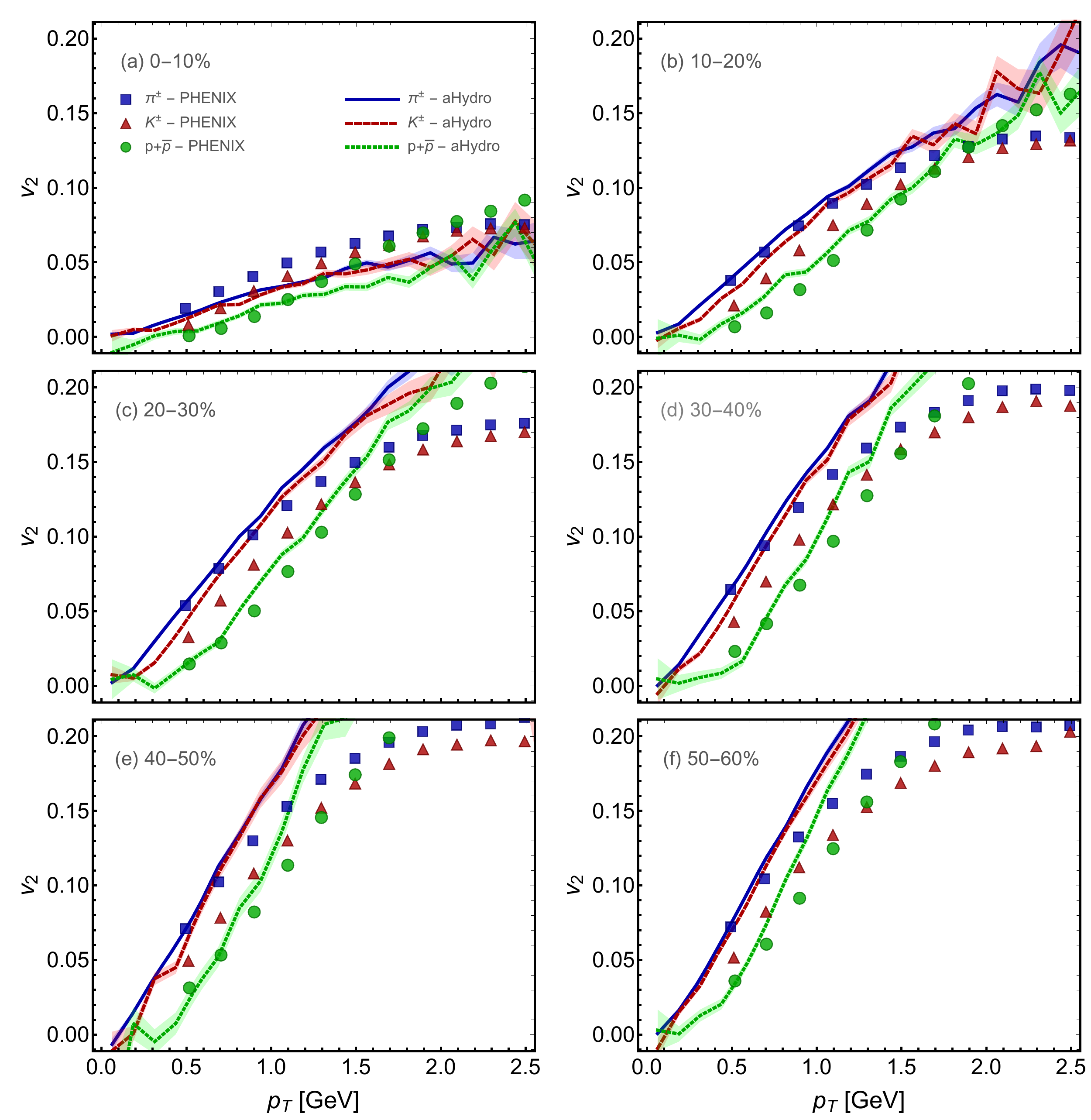}
}
\caption{aHydroQP results for the identified particle elliptic flow as a function of transverse momentum compared to experimental data from the PHENIX collaboration \cite{Adare:2014kci}.  The blue solid line, red dashed line, and green dotted lines are the aHydroQP predictions for pions, kaons, and protons, respectively.  The blue squares, red triangles, and green circles are experimental observations. The panels (a)-(f) show the centrality classes 0-10\%, 10-20\%, 20-30\%, 30-40\%, 40-50\%, and 50-60\%.  }
\label{fig:v2pt-combined}
\end{figure*}

\begin{figure*}[t!]
\centerline{
\includegraphics[width=.575\linewidth]{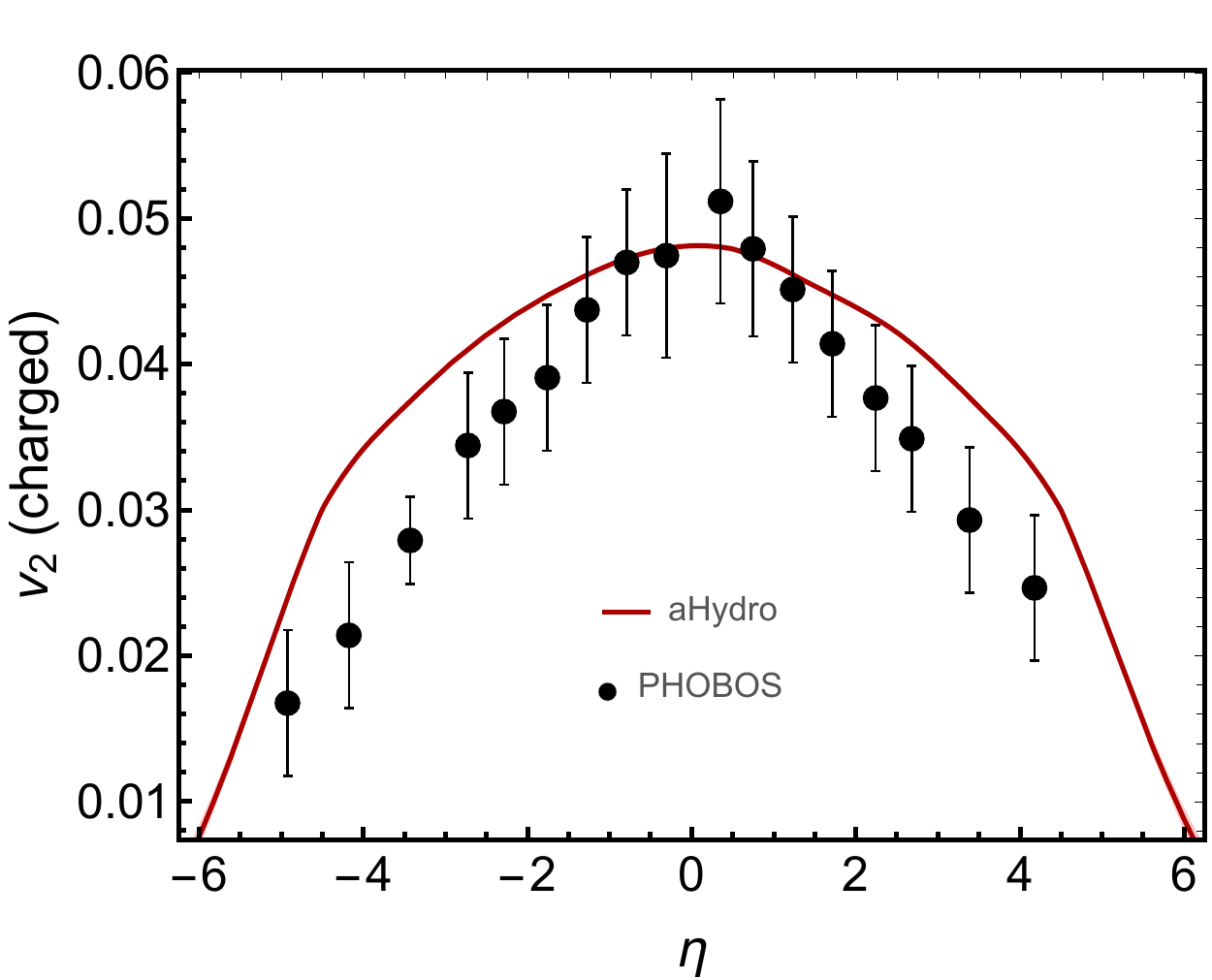}
}
\caption{Comparison between our aHydroQP model results for the integrated elliptic flow versus pseudorapidity and experimental data from the PHOBOS collaboration \cite{Back:2004zg,Back:2004mh}.  The aHydroQP results are indicated by a solid red line and the PHOBOS data by black points with error bars.  The experimental error bars include both statistical and systematic uncertainties.
}
\label{fig:v2eta-Charged}
\end{figure*}

In Fig.~\ref{fig:v2pt-combined} we present our results for the identified particle elliptic flow for pions, kaons, and protons as a function of transverse momentum.  We compare our model predictions with experimental observations available from the PHENIX collaboration \cite{Adare:2014kci}.  The blue solid line, red dashed line, and green dotted lines are the aHydroQP predictions for pions, kaons, and protons, respectively.  The blue squares, red triangles, and green circles are experimental observations.   As can be seen from this figure, the mass ordering observed is qualitatively reproduced, however, the agreement with the RHIC data is poorer than what was found when comparing to ALICE data from LHC 2.76 TeV collisions \cite{Alqahtani:2017jwl,Alqahtani:2017tnq}.  We also note that in all cases shown, the kaon elliptic flow is not well reproduced.  This is similar to what other modern viscous hydrodynamics codes coupled to hadronic afterburners have found when including bulk viscous effects at RHIC energies \cite{Ryu:2017qzn,hannahQM2018}.

Finally, in Fig.~\ref{fig:v2eta-Charged} we present a comparison between our aHydroQP model results for the integrated elliptic flow versus pseudorapidity and experimental data from the PHOBOS collaboration  \cite{Back:2004zg,Back:2004mh}.  The aHydroQP results are indicated by a solid red line and the PHOBOS data by black points with error bars.  The experimental error bars include both statistical and systematic uncertainties.  As can be seen from this figure, aHydroQP does a quite good job in reproducing the overall shape of the elliptic flow versus pseudorapidity, however, the aHydroQP model results have a slightly wider profile than the experimental observations.  This should be contrasted with aHydroQP applied to LHC 2.76 TeV energies where it was found that the elliptic flow dropped off too slowly to accurately describe the experimental observations of the ALICE collaboration \cite{Alqahtani:2017tnq}.  Explanations for the discrepancy could include our assumption of a temperature-independent shear viscosity to entropy density ratio or event plane decorrelation in the rapidity direction.  We plan to investigate these effects in the context of aHydroQP in a forthcoming publication.

\section{Conclusions and outlook}
\label{sec:conclusions}

In this paper we have applied the aHydroQP formalism to phenomenological predictions for a variety of observables measured at RHIC highest beam energies.  We presented comparisons of identified particle spectra in different centrality classes, charged particle multiplicity versus pseudorapidity, identified particle multiplicity versus centrality across a wide range of particle species, identified particle elliptic flow versus transverse momentum, and charged particle elliptic flow as a function of transverse momentum and rapidity.  We used the same aHydroQP and hadronic production/feed down codes that were used previously to describe LHC 2.76 TeV data.  

The aHydroQP hydrodynamic model used includes effects of both shear and bulk viscosities in addition to an infinite number of transport coefficients computed self-consistently in the relaxation time approximation.   To convert to the final state hadrons, we used anisotropic Cooper-Frye freeze-out performed on a fixed-energy-density hypersurface and computed the production/feed down using a customized version of Therminator 2. We found good agreement with many heavy-ion collision observables using only smooth Glauber initial conditions parameterized by an initial central temperature of $T_0 = 455$ MeV, a constant shear viscosity to entropy density ratio $\eta/s= 0.179$, and a switching (freeze-out) temperature of $T_{\rm FO}=130$ MeV.

Looking forward, the next hurdles on the aHydroQP front are to include the off-diagonal momentum-space anisotropies in the calculation and to produce results with fluctuating initial conditions.  In addition, we are currently studying the effects of different parameterizations of the temperature dependence of $\eta/s$.  The results presented here should provide a baseline for these future works.

\acknowledgments

D. Almaalol was supported by a fellowship from the University of Zawia, Libya. M. Alqahtani was supported by Imam Abdulrahman Bin Faisal University, Saudi Arabia.  M. Strickland was supported by the U.S. Department of Energy, Office of Science, Office of Nuclear Physics under Award No. DE-SC0013470.

\bibliography{rhic}

\end{document}